\documentclass[aps,prb,twocolumn,superscriptaddress,bibliography]{revtex4-2}
\usepackage{amsmath}
\usepackage{amssymb}
\usepackage{amsfonts}
\usepackage{bm}
\usepackage{mathrsfs}
\usepackage{wasysym}
\usepackage{esint}
\usepackage{dcolumn}
\usepackage{longtable}
\usepackage{multirow}
\usepackage{xspace}
\usepackage{graphicx}
\usepackage{xcolor}
\usepackage{microtype}
\usepackage[colorlinks=true, letterpaper=true, pdfstartview=FitV, linkcolor=blue, citecolor=blue, urlcolor=blue]{hyperref}
\usepackage{braket}

\makeatletter
\providecommand{\tabularnewline}{\\}
\makeatother

\begin{document}
\title{Valley Berry curvature dipole induced nonlinear valley Hall effect}

\author{Lulu Xiong}
\affiliation{Institute of Applied Physics and Materials Engineering, Faculty of Science and Technology, University of Macau, Taipa, Macau SAR, China}

\author{Xue-Jin Zhang}
\affiliation{Institute of Applied Physics and Materials Engineering, Faculty of Science and Technology, University of Macau, Taipa, Macau SAR, China}

\author{Zeying Zhang}
\affiliation{College of Mathematics and Physics, Beijing University of Chemical Technology, Beijing 100029, China}

\author{Yinning Zhou}
\affiliation{Institute of Applied Physics and Materials Engineering, Faculty of Science and Technology, University of Macau, Taipa, Macau SAR, China}

\author{Jin Cao}
\email{caojin.phy@gmail.com}
\affiliation{Research Laboratory for Quantum Materials, Department of Physics and Materials, The Hong Kong Polytechnic University, Hong Kong SAR, China}

\author{Cong Xiao}
\affiliation{Interdisciplinary Center for Theoretical Physics and Information Sciences (ICTPIS), Fudan University, Shanghai 200433, China}

\author{Shengyuan A. Yang}
\affiliation{Research Laboratory for Quantum Materials, Department of Physics and Materials, The Hong Kong Polytechnic University, Hong Kong SAR, China}

\begin{abstract}
Valley Hall effect is a signature effect in the field of valleytronics, and recent studies have pushed this effect into the nonlinear regime. Here, we reveal a previously unexplored type of nonlinear valley Hall effect which arises from a valley Berry curvature dipole (vBCD) mechanism. We show this nonlinear valley Hall effect is forbidden for conventional time-reversal-connected valleys, but is supported in the class of valleytronic systems featuring time-reversal-invariant valleys. The candidate layer groups and detailed symmetry constraints on vBCD are obtained. It shows the nonlinear valley Hall response, as well as the nonlinear charge Hall response, can be well controlled by tuning the driving field direction. We demonstrate our proposal in an effective model study and in a concrete material example, strained Nb$_{3}$SBr$_{7}$, by first-principles calculations. The nonlocal transport signature of this vBCD induced nonlinear valley Hall effect is also discussed. 
\end{abstract}

\maketitle

\section{Introduction}

Valleytronics aims to utilize the electrons' valley degree of freedom, associated with the local extrema of conduction or valence band in momentum space, for efficient information processing~\cite{Gunawan2006Valley,Rycerz2007Valley,Xiao2007Valley,Yao2008Valley,Yao2009Edge,Xiao2012Coupled,Cai2013Magnetic,Xu2014Spin,Mak2014valley,Gorbachev2014Detecting,Sui2015Gate,Shimazaki2015Generation,Pan2015Perfect,Pan2015Valley,Schaibley2016Valleytronics,Lee2016Electrical,Mak2018Lightvalley,Wu2019Intrinsic,Hung2019Direct,Yu2020Valley,Li2020Room,Jiang2022room,Luo2024Valleytronics}. A key effect in valleytronics is the valley Hall effect (VHE), where a valley current $\bm j^v$ is generated in a direction transverse to the driving electric field~\cite{Xiao2007Valley}.
So far, studies on VHE have been mainly on a class of two-dimensional (2D) hexagonal lattice materials, such as graphene and
H-phase transition metal dichalcogenides~\cite{Xiao2007Valley,Yao2008Valley,Mak2014valley,Schaibley2016Valleytronics}. In these materials, there is a pair of valleys, $K$ and $K'$, related by the time-reversal symmetry ($\mathcal T$). Here, the system as a whole preserves $\mathcal T$, but an individual valley does not. The VHE in such a system originates from mechanisms that are odd under $\mathcal T$. For example, it contains an important intrinsic mechanism from the valley-contrasting Berry curvature~\cite{Xiao2007Valley}.

Recently, inspired by the development in nonlinear Hall effects~\cite{Gao2014Field,Sodemann2015Quantum,Ma2018Observation,Kang2019Nonlinear,Du2019Disorder,xiao2019theory,Wang2021NAHE,Liu2021NAHE,Wang2023Quantum,Gao2023Quantum,Huang2025Scaling,Xiao2025Proper}, nonlinear valley transport has been attracting great interest~\cite{Yu2014Nonlinear,Rodin2016Valley,Das2024Nonlinear,Cao2025Nonlocal,Zhou2025Nonlinear,He2026Observation,Jiang2026Tunable,Wan2024Strongly,Zhang2025Intrinsic,Sharma2025Nonlinear,Wan2025Extrinsic,Wu2025Intrinsic}. Particularly, nonlinear VHE, with $j^v\propto E^2$, has been proposed in those 2D hexagonal lattice valleytronic materials. Like linear VHE, the nonlinear VHE in such systems must also originate from $\mathcal T$-odd mechanisms. For example, Yu \emph{et al.}~\cite{Yu2014Nonlinear} proposed a Drude-like mechanism, arising from the second-order field correction to the electron distribution. There also exists an intrinsic mechanism, associated with the band geometric quantity of Berry-connection polarizability~\cite{Gao2014Field,Liu2022Berry,Das2024Nonlinear}. The theory of nonlocal transport from linear and nonlinear VHEs has also been developed.
Interestingly, it was found that unlike linear VHE, the direct and inverse processes of nonlinear VHE are not reciprocal, and they involve distinct mechanisms~\cite{Cao2025Nonlocal}. This prediction and the proposed scaling law for nonlocal signals from nonlinear VHE were successfully confirmed in a recent experiment on a graphene superlattice system~\cite{He2026Observation}.

From symmetry point of view, the constraint that VHEs (both linear and nonlinear) must involve $\mathcal T$-odd mechanisms is a result of the valley character of those canonical valleytronic materials, i.e., their valleys are $\mathcal T$-connected. Recently, Ref.~\cite{Cao2026Eccentricity} proposed a new class of valleytronic systems, featuring time-reversal-invariant valleys (TRIVs). Here, each valley is sitting at a $\mathcal T$-invariant momentum point, and the two valleys are not connected by $\mathcal T$ but by some crystal symmetry. This fundamentally changes the symmetry character of valley physics, generating new effects.
For example, it was shown that TRIVs host an unconventional linear VHE depending on the eccentricity of valley Fermi surface~\cite{Cao2026Eccentricity}.
Motivated by this development, one may naturally wonder: Going beyond the conventional $\mathcal T$-connected valleys, does nonlinear VHE also contain new physics?

In this work, we answer this question by proposing
a valley Berry curvature dipole (vBCD) induced nonlinear VHE in TRIV systems. Distinct from previously proposed nonlinear VHEs, this vBCD mechanism is of $\mathcal T$-even character. It is strictly forbidden for $\mathcal T$-connected valleys, but may be allowed for TRIVs.
Through symmetry analysis, we identify the layer groups that support this effect, which belong to the crystal class of centered rectangular lattices. The vBCD and (charge) BCD are always perpendicular to each other. Hence, one can selectively generate valley Hall current or charge Hall current, by controlling the driving field direction. Using first-principles calculations, we show that strained monolayer Nb$_{3}$SBr$_{7}$ can host a sizable vBCD, and the resulting nonlinear VHE conductivity is estimated.
Furthermore, we consider the nonlocal transport signal resulting from this vBCD-induced nonlinear VHE, and predict how the nonlocal voltage signals scale with resistivity, an important characteristic that can be directly tested in experiment.
These findings uncover a new mechanism of nonlinear valley transport, broaden the scope of valleytronics, and open a new route to harnessing valley degree of freedom for applications.

\begin{figure}
\centering{}\includegraphics[width=8.5cm]{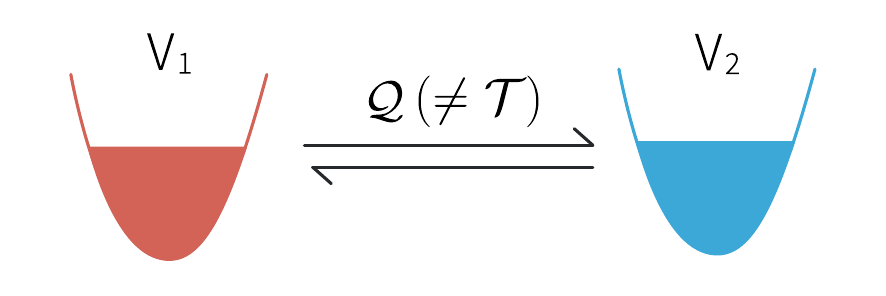}
\caption{\label{fig1} Schematic illustration of a valleytronic system with two valleys $V_1$ and $V_2$. For a TRIV system, the two valleys are connected by some crystal symmetry $\mathcal{Q}$, rather than by time-reversal symmetry $\mathcal{T}$.}
\end{figure}

\section{valley Berry curvature dipole}

Let's consider a 2D nonmagnetic system with two valleys in the conduction band, denoted as $V_{1}$ and $V_{2}$, and assume the Fermi level is near the band edge such that only the two valleys contribute to transport.
The case with valence band valleys can be treated in a similar way.
As shown in Ref.~\cite{Cao2026Eccentricity}, such systems can be classified into two classes. The first class is that the two valleys are related by $\mathcal T$. Evidently, each valley here cannot be at a $\mathcal T$-invariant momentum point.
The other class is with two valleys located at $\mathcal T$-invariant points, i.e., with TRIVs. Here, to ensure the energy degeneracy of $V_1$ and $V_2$, there should exist some crystal symmetry $\mathcal Q$ (other than $\mathcal T$) that connects the two valleys (see Fig.~\ref{fig1}).

\begin{figure}
\centering{}\includegraphics[width=8.5cm]{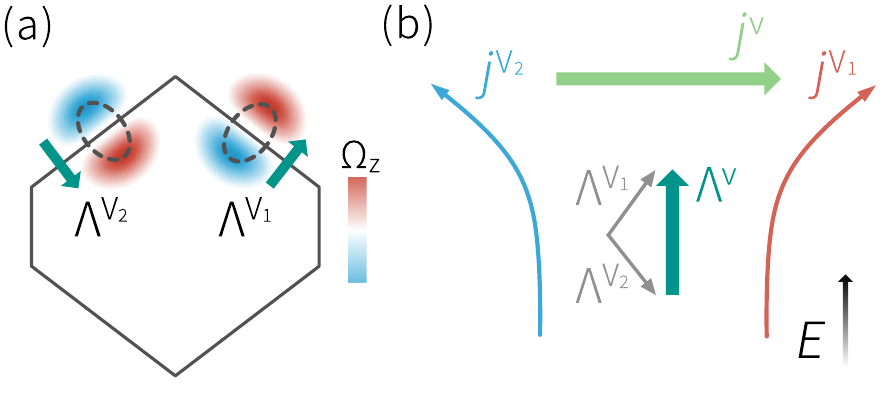}
\caption{\label{fig2}Schematic illustration of vBCD and the induced nonlinear VHE. (a) At each individual valley, the valley-resolved BCD ($\boldsymbol{\Lambda}^{V_1}$ and $\boldsymbol{\Lambda}^{V_2}$) arises from the asymmetric distribution of Berry curvature. (b) The valley-contrasting component of $\boldsymbol{\Lambda}^{V_1}$ and $\boldsymbol{\Lambda}^{V_2}$ gives rise to the vBCD. When an electric field is applied along the vBCD direction, a nonlinear valley Hall current $\boldsymbol{j}^{v}$ is produced in the transverse direction.}
\end{figure}

Consider the nonlinear current response from carriers in one of the valleys $V_i$ ($i=1,2$), driven by an electric field in the 2D plane. Sodemann and Fu~\cite{Sodemann2015Quantum} showed that
there exists a nonlinear Hall current from the BCD mechanism, which is of $\mathcal T$-even character. For valley $V_i$, this nonlinear current response is given by (take $e=\hbar=1$):
\begin{eqnarray}
\boldsymbol{j}^{V_i}=-\tau\left(\boldsymbol{\Lambda}^{V_i}\cdot\boldsymbol{E}\right)\hat{z}\times\boldsymbol{E},
\end{eqnarray}
where $\tau$ is the carrier relaxation time, $\hat z$ is the unit vector normal to the 2D plane, and $\boldsymbol{\Lambda}^{V_i}$ is the BCD of carriers in valley $V_i$. $\boldsymbol{\Lambda}^{V_i}$ can be expressed as
\begin{eqnarray}\label{LVi}
\Lambda_{a}^{V_i}=-\int_{V_{i}}\left[d\boldsymbol{k}\right]f_{0}^{\prime}(v_a)_{n\bm k}\Omega_{n\bm k},
\end{eqnarray}
where subscript $a$ denotes the two in-plane Cartesian components $x$ and $y$, $\left[d\boldsymbol{k}\right]$ is a shorthand notation for $\sum_{n}d\boldsymbol{k}/\left(2\pi\right)^{2}$, $n$ is the band index, $f_{0}$ is the Fermi-Dirac distribution function, $(v_a)_{n\bm k}$ is the $a$-component of band velocity of state $\ket{u_{n\bm k}}$, $\Omega$ is the $z$-component of Berry curvature
\begin{eqnarray}
  \Omega_{n\bm k}=-2\text{Im}\braket{\partial_{k_x} u_{n\bm k}|\partial_{k_y} u_{n\bm k}},
\end{eqnarray}
and the integral is over states of $V_{i}$ valley. This valley-resolved BCD $\boldsymbol{\Lambda}^{V_i}$
characterizes the asymmetric distribution of Berry curvature in valley $V_i$, as schematically illustrated in Fig.~\ref{fig2}(a).

Combining the two contributions $\boldsymbol{j}^{V_1}$ and $\boldsymbol{j}^{V_2}$, one obtains the total nonlinear charge Hall current from BCD, given by
\begin{eqnarray}
\boldsymbol{j}^{c}=\boldsymbol{j}^{V_1}+\boldsymbol{j}^{V_2}=-\tau\left(\boldsymbol{\Lambda}^{c}\cdot\boldsymbol{E}\right)\hat{z}\times\boldsymbol{E},
\end{eqnarray}
where $\boldsymbol\Lambda^{c}=\boldsymbol\Lambda^{V_1}+\boldsymbol\Lambda^{V_2}$ is the total (charge) BCD of the system.

Meanwhile, we are more interested in the nonlinear valley Hall current, which is the difference between $\boldsymbol{j}^{V_1}$ and $\boldsymbol{j}^{V_2}$:
\begin{eqnarray}\label{jv}
\boldsymbol{j}^{v}=\boldsymbol{j}^{V_1}-\boldsymbol{j}^{V_2}=-\tau\left(\boldsymbol{\Lambda}^{v}\cdot\boldsymbol{E}\right)\hat{z}\times\boldsymbol{E},
\end{eqnarray}
where $\boldsymbol{\Lambda}^{v}$ is what we define as vBCD, given by
\begin{eqnarray}
  \boldsymbol{\Lambda}^{v}\equiv \boldsymbol\Lambda^{V_1}-\boldsymbol\Lambda^{V_2}.
\end{eqnarray}
These are the formulas describing nonlinear VHE from vBCD mechanism.

One can make the following observations. First, from Eq.~(\ref{jv}), the nonlinear valley Hall current
has a simple  dependence on the direction of driving $E$ field. It depends on the cosine of the angle between vBCD $\boldsymbol{\Lambda}^{v}$ (which is a vector in the 2D plane) and $E$ field. The magnitude $j^v$ is maximal when $\bm E$ is parallel (or antiparallel) to vBCD. And the response vanishes when  $\bm E$ is perpendicular to vBCD. For the case shown in Fig.~\ref{fig2},
$\boldsymbol{\Lambda}^{v}$ vector is in the $y$ direction, so the nonlinear VHE response is strongest when $E$ field is also in $y$ (or $-y$) direction, as illustrated in Fig.~\ref{fig2}(b).

Second, this nonlinear VHE in Eq.~(\ref{jv}) is proportional to the relaxation time $\tau$. This  is distinct from the nonlinear Drude mechanism which is proportional to $\tau^2$~\cite{Yu2014Nonlinear}; and is also distinct from the intrinsic mechanism as well as so-called zeroth-order extrinsic mechanisms~\cite{Gao2014Field,Wang2021NAHE,Liu2021NAHE,Das2024Nonlinear,Huang2025Scaling}, which are independent of $\tau$. Indeed, as we mentioned, those mechanisms are of $\mathcal T$-odd character, whereas our vBCD mechanism is of $\mathcal T$-even character. They are fundamentally different.

Third, we mentioned that there are two classes of valleytronic systems, according to the constraint by $\mathcal T$ symmetry. Then, which class can support this vBCD induced nonlinear VHE? Since this response is even under $\mathcal T$, for conventional valleytronic systems with $\mathcal T$-connected valleys, we must have $\boldsymbol\Lambda^{V_1}=\boldsymbol\Lambda^{V_2}$, which dictates a vanishing vBCD. Therefore, this effect cannot happen in those conventional systems.
On the other hand, TRIV systems do not suffer from this constraint, so they may support a finite vBCD.
In other words, our proposed vBCD induced nonlinear VHE may be realized in TRIV systems.
In the next section, we shall perform a more detailed analysis to elucidate the symmetry conditions.

\section{Symmetry requirement}\label{sym}

We have seen that vBCD induced nonlinear VHE cannot be realized for conventional $\mathcal T$-connected valleys. It must require TRIV systems. Note that this is a necessary but not a sufficient condition. Other crystalline symmetries also impose constraints on this effect. Here, we shall analyze what TRIV systems can support vBCD induced nonlinear VHE.

The symmetry group $G$ of a valleytronic system can be decomposed into the following form:
\begin{eqnarray}
  G=G_V\amalg \mathcal{Q}G_{V},
\end{eqnarray}
where $G_V$ is the valley little group, i.e., the subgroup of $G$ which preserves each valley, and the set $\mathcal{Q}G_{V}$ contains all the symmetry elements that switch the two valleys. For TRIV systems, we have $\mathcal T\in G_V$.

One can see that (i) the form of valley-resulted BCDs $\boldsymbol{\Lambda}^{V_i}$ is constrained by $G_V$; (ii) besides $G_V$,
the form of vBCD is constrained further by $\mathcal{Q}G_{V}$. Specifically, according to its definition,  the transformation rule of vBCD $\boldsymbol{\Lambda}^{v}$ under a symmetry $\mathcal O\in G$ is given by
\begin{equation}
\Lambda^{v}_{a}=\epsilon_{v}\text{det}(\mathcal{O})\mathcal{O}_{zz}\mathcal{O}_{aa^{\prime}}\Lambda^{v}_{a^{\prime}},
\end{equation}
where $\epsilon_{v}=+1$ $(-1)$ for $\mathcal O\in G_V$ $(\in \mathcal{Q}G_{V})$, and the appearance of $\mathcal{O}_{zz}$ is because of the dependence on the $z$-component of Berry curvature in vBCD (see Eq.~(\ref{LVi})).

Using the above condition, we search through all 80 layer groups (LGs) that  support TRIVs to look for candidates that allow a nonzero vBCD. (The list of TRIV LGs was presented in Ref.~\cite{Cao2026Eccentricity}) We find that there are four candidate groups that can support vBCD induced nonlinear VHE. The results are summarized in Table~\ref{tab1}.

From Table~\ref{tab1}, one can see that all the candidate groups belong to the centered rectangular lattice class.
The TRIVs must be located at $S$ and $S'$ points of the Brillouin zone. For LG~10, the two TRIVs are connected by $C_{2x}$ symmetry, and vBCD is along this twofold axis (i.e., $x$ axis). In comparison, for LGs~13, 35, and 36, the two valleys are connected by mirror $M_x$, and vBCD is along this mirror line (i.e., the $y$ axis).

In addition, one can also easily analyze the symmetry constraints on (charge) BCD $\boldsymbol\Lambda^{c}$ in these LGs.
Interestingly, one finds that the direction of $\boldsymbol\Lambda^{c}$ turns out to be always perpendicular to $\boldsymbol\Lambda^{v}$, as shown in Table~\ref{tab1}. This is not a coincidence, since
\begin{equation}
  \boldsymbol\Lambda^{v}\cdot \boldsymbol\Lambda^{c}=(\Lambda^{V_1})^2-(\Lambda^{V_2})^2.
\end{equation}
For two valleys connected by some symmetry, the magnitudes $\Lambda^{V_i}$ of valley-resolved BCDs must equal, so $\boldsymbol\Lambda^{v}$ and $\boldsymbol\Lambda^{c}$ should be perpendicular. This feature offers a route for controlling the generation of Hall response. For example, in LG~10, by orienting $E$ field along the $x$ direction, i.e., along vBCD, one generates the maximal nonlinear VHE but the nonlinear charge Hall response vanishes. On the other hand, if rotating $E$ field to the $y$ direction, the nonlinear charge Hall response becomes maximal but nonlinear VHE is suppressed.

\begin{table}
\caption{\label{tab1}Layer groups that support vBCD induced nonlinear VHE. The TRIVs $S$ and $S^{\prime}$ in centered rectangular lattice are illustrated in the figure. $\mathcal{Q}$ denotes the symmetry that relates $S$ and $S^{\prime}$. The symmetry allowed component of vBCD and (charge) BCD are given in the last two columns. }
\begin{ruledtabular}
\renewcommand{\arraystretch}{1.5}
\begin{centering}
{\footnotesize{}%
\begin{tabular}{@{}cccc@{}c@{}}
TRIVs & $\mathcal{Q}$ & Generators of $G_{V}$ & vBCD & BCD\tabularnewline
\hline
\multirow{2}{*}{{\footnotesize\includegraphics[width=1.6cm]{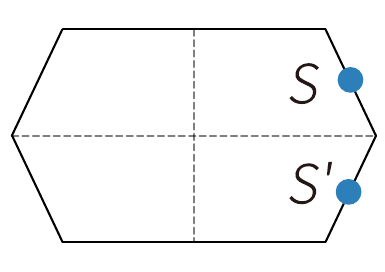}}} & $C_{2x}$  & \{$E$\} (LG 10) & $\Lambda^{v}_{x}$ & $\Lambda^{c}_{y}$\tabularnewline
 & $M_{x}$  & \{$E$\} (13); \{$M_{z}$\} (35, 36) & $\Lambda^{v}_{y}$ & $\Lambda^{c}_{x}$\tabularnewline
\end{tabular}}{\footnotesize\par}
\par\end{centering}
\end{ruledtabular}
\end{table}

\section{Model study}

\begin{figure}[t]
\centering{}\includegraphics[width=8cm]{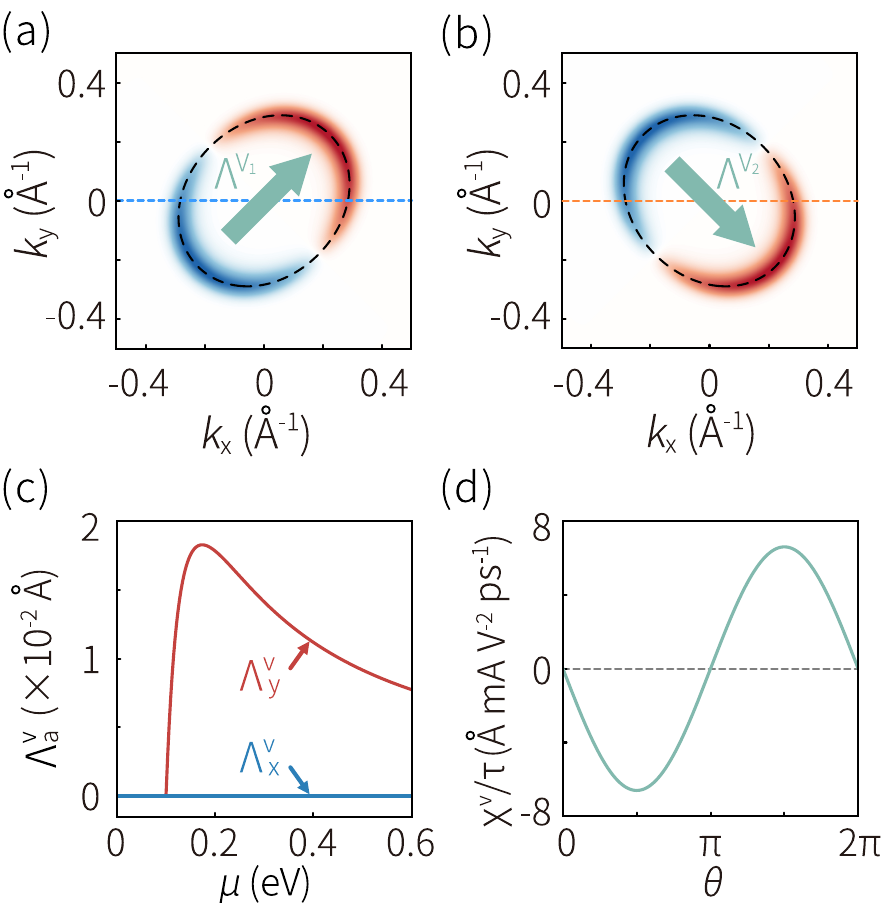}
\caption{\label{fig3}Exhibition of vBCD and the induced nonlinear VHE in an effective model of LG 13. (a,b) The Berry curvature distributions of $V_1$ and $V_2$ valleys on the Fermi surface. The dashed lines represent the Fermi surfaces at a chemical potential of $\mu=0.2$ eV. The arrows indicate the direction of valley-resolved BCD. (c) The $x$- and $y$-components of vBCD as functions of chemical potential. (d) The calculated nonlinear valley Hall conductivity as a function of the direction of electric field.}
\end{figure}

We first demonstrate our proposed vBCD induced nonlinear VHE in a TRIV effective model. We consider a $k\cdot p$ model for
LG~13 ($cm111^{\prime}$). From Table~\ref{tab1}, the two TRIVs are located at  $S$ and $S^{\prime}$ points.
Here, the valley little group $G_V$ contains only $\mathcal T$, and $\mathcal Q=M_x$ connects the two valleys. As before, assume the two TRIVs are for the conduction band of a semiconductor band structure, and we consider electron doped case.
Nevertheless, since Berry curvature embodies an interband coherence, to describe vBCD, the model must contain at least two bands. Hence, we need also include the valence band states around $S$ and $S^{\prime}$ in the model.

We construct the following $k\cdot p$ model for the two TRIVs $V_i$ ($i=1,2$):
\begin{eqnarray}\label{kp}
\mathcal{H}_{V_{i}}=\left(\Delta+\gamma k^{2}\right)\sigma_{x}+\left(\xi_{i}\nu_x k_{x}-\nu_y k_{y}\right)\sigma_{y}+\lambda\sigma_{z}.
\end{eqnarray}
Here, the momentum $\bm k$ is measured from the valley center $S$ (or $S^{\prime}$),
$\sigma$'s are the Pauli matrices corresponding to the two bands, $\xi_{i}=\pm1$ for $V_1$ and $V_2$ valleys respectively,
$\nu_x$, $\nu_y$, $\gamma$, $\Delta$, and $\lambda$ are real model parameters. Note that Eq.~(\ref{kp}) is not the most general model constrained by LG~13. For example, the symmetry-allowed $k^2$ term should in general be anisotropic. Here, for simplicity, we take it to be isotropic, because this does not affect the essential feature of vBCD, which is our focus here.

The spectrum and the eigenstates of this model can be easily obtained. Focus on the conduction band, which has energy dispersion of
\begin{eqnarray}
  \mathcal E_{V_i,c}=\Big[(\Delta+\gamma k^{2})^{2}+(\xi_{i}\nu_{x}k_{x}-\nu_{y}k_{y})^{2}+\lambda^{2}\Big]^{1/2}.
\end{eqnarray}
Taking $\lambda\ll \Delta$, near a valley center ($S$ or $S'$), its Berry curvature distribution is found to be
\begin{eqnarray}
\Omega_{V_i,c}\approx\frac{ \lambda\gamma}{\Delta^3}\left(\xi_{i}\nu_{x}k_{y}+\nu_{y}k_{x}\right).
\end{eqnarray}
In Fig.~\ref{fig3}(a) and ~\ref{fig3}(b), we plot the numerical result of Berry curvature distribution on a Fermi surface at $\mu=0.2$ eV (the band edge here is at $\mu=0.1$ eV), which indeed show an asymmetric dipole like pattern.
And the direction of dipole differs between the two valleys (the two dipoles are related by $M_x$), leading to a nonzero vBCD. From Fig.~\ref{fig3}(c), one can easily see that the vBCD should be along the $y$ direction, consistent with our symmetry analysis in Table~\ref{tab1}.

For chemical potential $\mu$ near the band edge with  $0<\mu-\Delta\ll \Delta$, one finds that
\begin{equation}
\boldsymbol{\Lambda}^{v}\approx\frac{\lambda\nu_{x}\left(\mu-\Delta\right)}{2\pi\mu^{3}}\hat{y}.
\end{equation}
Meanwhile, the charge BCD is along the $x$ direction, given by
\begin{equation}
\boldsymbol{\Lambda}^{c}\approx\frac{\lambda\nu_{y}\left(\mu-\Delta\right)}{2\pi\mu^{3}}\hat{x}.
\end{equation}

For a driving $E$ field $\bm E=E(\cos\theta,\sin\theta)$ in the 2D plane, the vBCD induced nonlinear VHE leads to a valley current in the direction of $\hat z\times \bm E$, i.e., along $\hat n=(-\sin\theta, \cos\theta)$. Hence, one may write
\begin{eqnarray}
  \bm j^v=\chi^v E^2 \hat n,
\end{eqnarray}
with $\chi^v$ being the nonlinear valley Hall conductivity.
For the current case, $\chi^v$ is given by
\begin{eqnarray}\label{chiv}
  \chi^v=-\tau\Lambda^v \sin\theta,
\end{eqnarray}
which has a simple sine dependence on the direction of driving field. (Similarly, one can define the nonlinear charge Hall conductivity $\chi^c$, which has a $\cos\theta$ dependence.) The angular dependence of calculated $\chi^v$ is shown in Fig.~\ref{fig3}(d).

\section{Material example}

\begin{figure}
\centering{}\includegraphics[width=8.5cm]{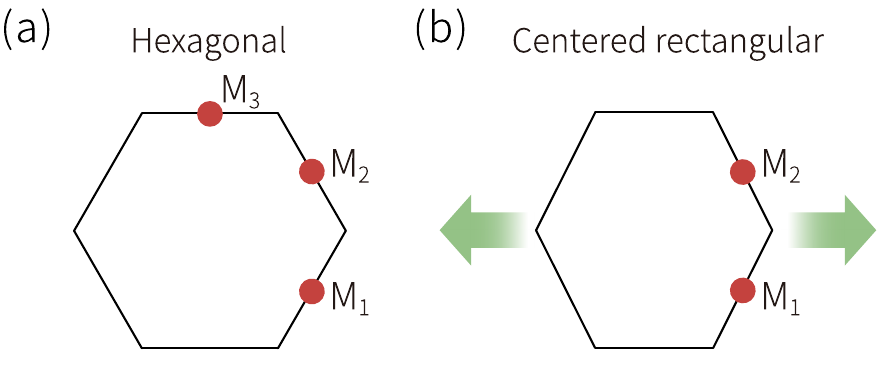}
\caption{\label{fig4}(a) Illustration of a hexagonal crystal having three $C_{3z}$-connected valleys at the three $M$ points. (b) Under a uniaxial strain, the hexagonal lattice becomes a centered rectangular lattice. And the energy degeneracy among the three $M$ valleys is lifted, resulting in a pair of TRIVs. }
\end{figure}

\begin{figure}
\centering{}\includegraphics[width=8.5cm]{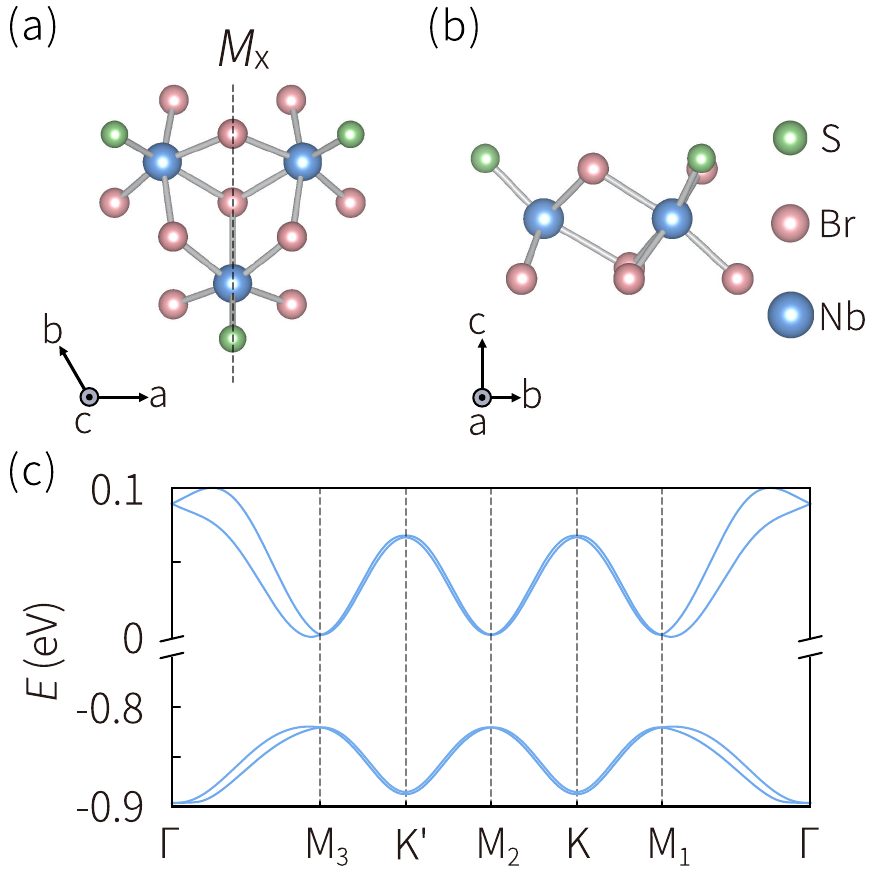}
\caption{\label{fig5}(a,b) Top and side views of monolayer Nb$_{3}$SBr$_{7}$. The mirror line associated with the $M_x$ symmetry is shown by the dashed line. (c) Calculated band structure of the unstrained monolayer Nb$_{3}$SBr$_{7}$.}
\end{figure}

The symmetry conditions obtained in Sec.~\ref{sym} offer useful guidance for the search of suitable material candidates.
We note that a centered rectangular lattice can be obtained from a hexagonal lattice by applying  uniaxial strain.
As illustrated in Fig.~\ref{fig4}(a), if a hexagonal lattice system hosts three $C_{3z}$ connected valleys at its $M$ points, then a uniaxial strain can lift the valley degeneracy and produce a pair of TRIVs for the resulting centered rectangular lattice (see Fig.~\ref{fig4}(b)).
Since strain engineering is particularly convenient for 2D materials, this idea greatly broadens the material systems for realizing our proposed physics.

Based on this idea, we propose the
monolayer Janus semiconductor Nb$_{3}$SBr$_{7}$ as a candidate material. Bulk Nb$_{3}$SBr$_{7}$ has been experimentally synthesized~\cite{Niobium1995}. It is a van der Waals layered material, so its 2D layers can be readily exfoliated.
The monolayer structure of Nb$_{3}$SBr$_{7}$ is shown in Fig.~\ref{fig5}(a) and \ref{fig5}(b). It consists of three atomic layers. The top layer contains both S and Br atoms, whereas the bottom layer contains only Br atoms, forming a Janus type structure. This strong structural asymmetry results in a non-centrosymmetric structure, which is needed for a nonzero vBCD.

For the unstrained monolayer, its LG is $p3m11^{\prime}$ (LG 69), which preserves both $C_{3z}$ and $M_{x}$ symmetries. Using first-principles calculations, we find that its optimized lattice constant is $7.17\,\text{\AA}$. This value is close to the experimental lattice constant reported for the bulk material~\cite{Niobium1995}. The calculated band structure is shown in Fig.~\ref{fig5}(c), exhibiting a semiconducting character with a band gap of approximately $0.82~\mathrm{eV}$. The conduction band minimum and the valence band maximum are found to be near the three $C_{3z}$-connected $M$ points (denoted as $M_{1}$, $M_{2}$ and $M_{3}$ here), consistent with the previous studies~\cite{Nb3SBr72022Janus}.
The valley structure here conforms with the picture in Fig.~\ref{fig4}(a).

\begin{figure}[t]
\centering{}\includegraphics[width=8.5cm]{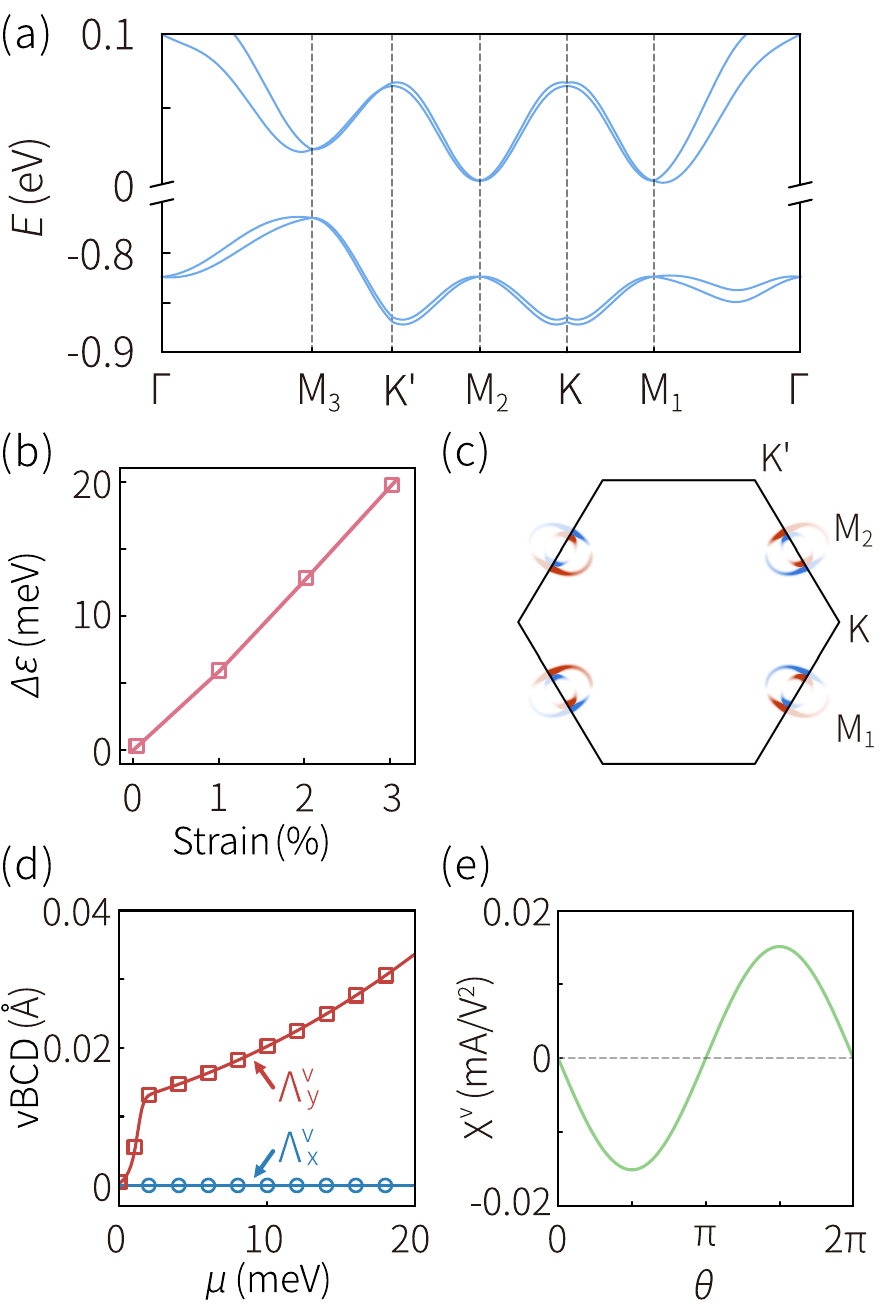}
\caption{\label{fig6}(a) Calculated band structure of monolayer Nb$_{3}$SBr$_{7}$ with $3\%$ uniaxial tensile strain along the $x$ direction. (b) Energy separation $\Delta \varepsilon=\varepsilon({M_3})-\varepsilon({M_1})$ (for the conduction band valleys) as a function of strain along $x$. (c) Distribution of Berry curvature for the band structure in (a) on the Fermi surface at $\mu=10$~meV. (d) Calculated vBCD as a function of chemical potential. (e) The vBCD induced nonlinear valley Hall conductivity as a function of the direction of driving field. Here, we take $\mu=10$~meV, $\tau=0.01$~ps, and $T=2$~K.}
\end{figure}

We then consider the effects of uniaxial strain along the $x$ direction. Under strain, the LG of Nb$_{3}$SBr$_{7}$ monolayer reduces to $cm111^{\prime}$ (LG 13), where the $C_{3z}$ symmetry is broken but $M_{x}$ is still maintained. The uniaxial strain lifts the degeneracy among the three $M$ valleys. However, the remaining $M_{x}$ symmetry still enforces the degeneracy between valleys at $M_{1}$ and $M_{2}$. (Here, $M_{1}$ and $M_{2}$ just correspond to $S$ and $S'$ for a centered rectangular lattice. For easy understanding, we shall use $M_{1}$ and $M_{2}$, instead of $S$ and $S'$, in the discussion below.)

Figure ~\ref{fig6}(a) shows the calculated band structure under a $3\%$ uniaxial tensile strain along $x$. One can see that its
conduction band valleys indeed split, and the new band edge consists of valleys at $M_1$ and $M_2$.
The energy splitting $\Delta\varepsilon$ between $M_3$ and $M_1/M_2$ is about 20~meV. In Fig.~\ref{fig6}(b), we further plot the calculated
energy splitting $\Delta\varepsilon$ versus the applied strain, which exhibits an approximately linear dependence at small strains.
Meanwhile, in Fig.~\ref{fig6}(a), the valence band splits in the opposite manner, with $M_3$ valley being the new band edge.
Below, we shall focus on the conduction band valleys, which provide the required TRIV system as in Fig.~\ref{fig2}(a).

Taking the band structure in Fig.~\ref{fig6}(a) (i.e., under $3\%$ strain), Figure \ref{fig6}(c) plots the Berry curvature on the Fermi surface at $\mu=10$~meV (we set zero energy at the conduction band minimum). One observes a sizable and asymmetric Berry curvature distribution, giving rise to a finite valley-resolved BCD.
The distribution patterns at the two valleys are connected by the mirror $M_x$, leading to a vBCD along $y$.

The calculated vBCD versus chemical potential $\mu$ is plotted in Fig.~\ref{fig6}(d). The result confirms that vBCD is along the $y$ direction. Its value reaches about $\sim 0.02~\text{\AA}$ at $\mu=10$ meV. The associated nonlinear valley Hall conductivity $\chi^v$ is given by Eq.~(\ref{chiv}).
Taking relaxation time $\tau \sim 0.01 $~ps, which is typical for 2D materials, we plot the calculated $\chi^v$ versus angle $\theta$ in Fig.~\ref{fig6}(e). The nonlinear VHE is strongest when the applied $E$ field is along the $y$ direction (i.e., along vBCD), where the magnitude of $\chi^v$ is $\sim 0.18$ $\mathrm{mA/V^2}$.

\section{Discussion and conclusion}

\begin{figure}
\centering{}\includegraphics[width=8.5cm]{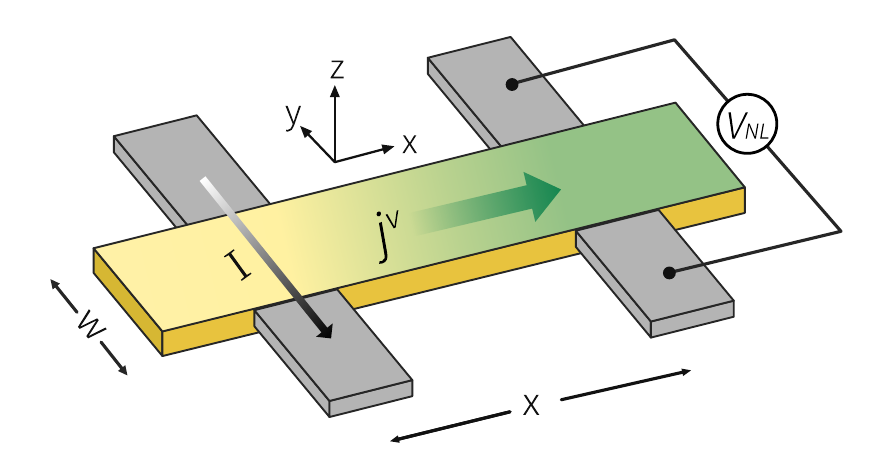}
\caption{\label{fig7} Schematic illustration of the nonlocal measurement setup.}
\end{figure}

Nonlocal measurement provides an electric approach for probing VHEs~\cite{Abanin2009Nonlocal,Gorbachev2014Detecting,Sui2015Gate,Shimazaki2015Generation,Beconcini2016Nonlocal,Wu2019Intrinsic,He2026Observation} (see Fig.~\ref{fig7}). For example, the recent experiment by He \emph{et al.}~\cite{He2026Observation} successfully detected the nonlinear VHE in a graphene superlattice with $\mathcal T$-connected valleys.
In such experiments, besides VHE, one of the key processes is the inverse VHE, which converts the valley current generated by VHE back into a charge current and hence leads to the nonlocal voltage signal. In the case of linear valley responses, the direct and inverse VHEs are reciprocal to each other and involve the same mechanism, in the spirit of Onsager reciprocity. However, for nonlinear valley responses, Ref.~\cite{Cao2025Nonlocal} showed that the two processes exhibit distinct symmetry properties and are not reciprocal to each other for $\mathcal T$-connected valleys.

As for the TRIV systems studied here, the inverse nonlinear VHE is described by~\cite{Cao2025Nonlocal}
\begin{equation}
\boldsymbol{j}^c=-\tau\hat{z}\times\boldsymbol{F}\left(\boldsymbol{\Lambda}^{c}\cdot\boldsymbol{F}\right),
\end{equation}
where $\bm{F}=-\nabla n^{v}/(2\nu)$ is the valley density gradient, acting as the driving force for the inverse process, $n^{v}$ is the valley density imbalance and $\nu$ is the density of states at Fermi level of one valley. Importantly, the geometric quantity that enters the inverse process is the charge BCD rather than the vBCD.

In experiment, the standard technique is to apply a low-frequency ac driving current $\tilde{I}=I\cos\omega t$, and to measure the nonlocal voltages $V_{\mathrm{NL}}$ of different harmonic channels using lock-in amplifier~\cite{Ma2018Observation,Kang2019Nonlinear,Lai2021Third}.
Following the approach in Ref.~\cite{Cao2025Nonlocal}, one finds that
the nonlinear VHE contributes to the nonlocal voltages in the third- and fourth-harmonic channels, with
\begin{eqnarray}
V^{3\omega}_{\mathrm{NL}}=\frac{I^{3}\rho^{6}}{4\pi\ell^{2}_{v}}\sigma^{v}\chi^{v}\zeta^{v}e^{-2x/\ell_{v}},\label{v3w}
\end{eqnarray}
and
\begin{eqnarray}
V^{4\omega}_{\mathrm{NL}}=\frac{I^{4}\rho^{7}}{8\pi^{2}w\ell^{2}_{v}}\left(\chi^{v}\right)^{2}\zeta^{v}e^{-2x/\ell_{v}}.\label{v4w}
\end{eqnarray}
Here, $w$ is the sample width, $\ell_{v}$ is the valley diffusion length, $\rho=\sigma^{-1}$ is the linear resistivity, $\sigma^{v}$ is the linear valley Hall conductivity, $\chi^{v}$ and $\zeta^{v}$ are conductivities of the direct and inverse nonlinear VHEs, respectively. It should be noted that the formulas (\ref{v3w}) and (\ref{v4w}) are general, applying to both
$\mathcal T$-connected valleys and TRIVs.

For a TRIV system, one has $\sigma^{v}\propto\tau$, corresponding to the eccentricity VHE proposed in Ref.~\cite{Cao2026Eccentricity}. If the direct and inverse nonlinear VHEs are dominated by vBCD and BCD mechanisms proposed here, one would have both $\chi^v, \zeta^v\propto \tau$. Therefore, the nonlocal signals would obey the following scaling behavior:
\begin{equation}
V^{3\omega}_{\mathrm{NL}}\propto\rho^{3},\qquad V^{4\omega}_{\mathrm{NL}}\propto\rho^{4}.
\end{equation}
This differs from the result for systems with $\mathcal T$-connected valleys~\cite{Cao2025Nonlocal}. For example, a scaling behavior of $V^{3\omega}_{\mathrm{NL}}\propto\rho^{5}$ has been observed in experiment~\cite{He2026Observation}.

It is worth noting that the scaling behavior of $\zeta^{v}$ can be independently determined through (local) nonlinear charge transport measurement~\cite{Du2019Disorder,xiao2019scaling}, based on which the scaling of nonlinear VHE $\chi^{v}$ can be investigated using Eqs.~(\ref{v3w}) and  (\ref{v4w}). Generally,
such scaling analysis provides a powerful tool for identifying the underlying mechanisms of nonlinear VHEs.

In conclusion, we have proposed a previously unexplored type of nonlinear VHE in TRIV systems. This effect is governed by a new band geometric quantity --- the vBCD. Via systematic symmetry analysis, we identify all LGs that support this effect. It is a general feature that the vBCD and charge BCD are perpendicular in the 2D plane, such that one can selectively drive the nonlinear Hall transport in either the valley or the charge channel. The key features of vBCD induced nonlinear VHE are demonstrated in a model study. And we also identify a suitable material candidate, the strained Nb$_{3}$SBr$_{7}$ monolayer, which exhibits a sizable nonlinear valley Hall conductivity. The scaling law for nonlocal measurement of our proposed effect is also discussed. By uncovering a novel mechanism of nonlinear valley transport, this work broadens the horizons of valleytronics and opens new avenues for utilizing the valley degree of freedom in technological applications.

\section*{Acknowledgments}
The authors thank W. T. Zhou and  D. L. Deng for discussions.
This work was supported by Quantum Science Strategic Special Project of Guangdong Province (No.~GDZX2504003) and Hong Kong PolyU start-up grant (P0057929). C.X. was supported by National Natural Science Foundation of China (Grant No.~12574114) and the startup funding from Fudan University.

\section*{Data availability}

The data that support the findings of this article are not publicly available. The data are available from the authors upon reasonable request.

\appendix

\section{Methods}

The first-principles calculations for monolayer Nb$_{3}$SBr$_{7}$ were performed using density functional theory as implemented in the VASP package~\cite{Kresse1993Ab,Kresse1996Efficiency,Kresse1996Efficient}. The projector augmented-wave method was employed with a plane-wave cutoff energy set to 500~eV~\cite{Bloechl1994Projector}. The Perdew-Burke-Ernzerhof functional~\cite{Perdew1996Generalized} was adopted to address the exchange-correlation energy.
The atomic positions were fully relaxed with a force tolerance of 0.01~eV$/\text{\AA}$. The spin-orbit coupling was included in calculations. The threshold for energy convergence was selected as 10$^{-6}$~eV.
For Brillouin zone sampling, a $\Gamma$-centered $k$-point mesh with size of $6\times6\times1$ was used.
The \textit{ab initio} tight-binding model was constructed using the Wannier90 package~\cite{Mostofi2014updated}, which was used for the calculations of band geometric quantities and the transport coefficients.

\bibliography{ref}

\end{document}